\documentclass[useAMS,usenatbib]{mnras}

\usepackage{graphicx}

\def\gsc{\gamma_{\rm sc}}
\def\gem{\gamma_{\rm em}}
\def\nuobs{\nu_{\rm obs}}
\def\nuem{\nu_{\rm em}}
\def\nucr{\nu_{\rm cr}}

\def\them{\theta_{\rm em}}

\def\rem{r_{\rm em}}

\def\esc{\eta_{\rm sc}}

\def\nct#1{\nocite{#1}}

\def\lap{\hbox{\hspace{4.3mm}}
         \raise1.5pt \vbox{\moveleft9pt\hbox{$<$}}
         \lower2.5pt \vbox{\moveleft9pt\hbox{$\sim$ }}
         \hbox{\hskip 0.02mm}}
\def\gap{\hbox{\hspace{4.3mm}}
         \raise1.5pt \vbox{\moveleft9pt\hbox{$>$}}
         \lower2.5pt \vbox{\moveleft9pt\hbox{$\sim$ }}
         \hbox{\hskip 0.02mm}}

\title[Evidence for scattering in radio pulsar profiles]{Geometric interpretation of radio pulsar profiles: 
evidence for intramagnetospheric scattering of curvature radiation}   
\title[Scattering as origin of profile morphology]{Intramagnetospheric 
scattering as the origin of radio pulsar profile
morphology}
\title{Geometric evidence for scattering as the origin of radio pulsar
profile morphology}
\title[Scattering of curvature radiation in pulsars]{Evidence for scattering of curvature radiation in radio pulsar
profiles}

\author[J.~Dyks]%, P.~Weltevrede, et al.]%, M.~Serylak, S.~Os{\l}owski, et al.]
{J.~Dyks%$^1$%, P. Weltevrede$^2$ and C.~Ilie$^2$
\\
Nicolaus Copernicus Astronomical Center, Polish Academy of Sciences, Rabia\'nska 8, 87-100, Toru\'n,
Poland\\
}
\begin{document}

%\date{Accepted .... Received 2015 August 10; in original form 2015 July 23}
\date{Accepted .... Received ...; in original form 2020 Sep 22}

%\pagerange{\pageref{firstpage}--\pageref{lastpage}} \pubyear{2002}

\maketitle

\label{firstpage}

\begin{abstract} 
Radio pulsars exhibit several  
unexplained phenomena, in particular the average pulse profiles with the apparent core-cone
structure and interesting frequency evolution.
I show that they can be interpreted through essential geometric properties of 
the inverse Compton scattering.  
If the scattering occurs in a dipolar magnetosphere and the mean-free-path
is long, a nested cone structure is expected with the cone size
ratio of two-thirds, which is consistent with observations. 
Being a discontinuous process, the scattering is consistent with the discrete
altitude structure of emission rings as derived from aberration-retardation effects.
Assuming that the upscattered signal is the curvature radiation (CR), one
can interpret the observed bifurcated components (BCs) as a magnified microbeam of
CR: the BCs are wide low-frequency CR microbeams that have been upshifted in frequency
with their width preserved by beam-copying scattering in divergent magnetic field. 
The large flux of BCs is partly caused by compression of the full emitted spectrum
into the narrow observed bandwidth, which explains why
the frequency-resolved BCs have the frequency-integrated shape.
The wide low-frequency microbeams can encompass 
large magnetospheric volumes, which considerably abates the requirements of the energy needed for
coherency. The properties of BCs thus suggest that the observed modulated radio flux is  strongly
affected by the scattering-driven blueshift and spectral compression. 
The relativistic beaming formula  ($1/\gamma$) is not always
applicable, in the sense that it may not be directly applied to some blueshifted
profile features.   
\end{abstract}


\def\lap{\hbox{\hspace{4.3mm}}
         \raise1.5pt \vbox{\moveleft9pt\hbox{$<$}}
         \lower1.5pt \vbox{\moveleft9pt\hbox{$\sim$ }}
         \hbox{\hskip 0.02mm}}

\def\rwobs{R_W}
\def\rwcon{R_W}
\def\rwstr{R_W}
\def\winobs{W_{\rm in}}
\def\woutobs{W_{\rm out}}
\def\phm{\phi_m}
\def\phmi{\phi_{m, i}}
\def\thm{\theta_m}
\def\dres{\Delta\phi_{\rm res}}
\def\win{W_{\rm in}}
\def\wout{W_{\rm out}}
\def\rin{\rho_{\rm in}}
\def\rout{\rho_{\rm out}}
\def\phin{\phi_{\rm in}}
\def\phout{\phi_{\rm out}}
\def\xin{x_{\rm in}}
\def\xout{x_{\rm out}}

\def\thmin{\theta_{\rm min}^{\thinspace m}}
\def\thmax{\theta_{\rm max}^{\thinspace m}}

%\begin{verse}
%\noindent 
%\it Putting together, inch by inch\\
% the starry worlds. From all the lost
%collections.\footnote{Adrienne Rich cited by Joanna M. Rankin (1983). 
%Fragment of ``For memory" from ``A wild patience has taken me this far", 
%Poems 1978-1981 by Adrienne Rich, W.W. Norton \& Company 1993. 
%}
%\end{verse}

\section{Introduction}

Despite more than half a century of observations, and vast data collected
for intensity profiles, polarization (Hankins \& Rankin 2010; Stinebring et
al.~1984; Noutsos et al.~2015; Tiburzi et al.~2013) \nct{hr10, scr84, nsk15,
tjb13} 
and flux density modulation (Deshpande \& Rankin 2001; Weltevrede et
al.~2007), \nct{wse2007, dr2001}
radiative pulsar properties resist most of interpretive efforts. The
curvature radiation (CR) continues to be regarded as a likely  
emission process (Gangadhara 2010; Gil, Lyubarskii \& Melikidze 2004; Luo \&
Melrose 1992; Mitra et al.~2009; Wang et al.~2015; Dyks et al.~2010,
hereafter DRD10), \nct{g10, glm04, lm92,
mgm2009, wwh15, drd10}
despite it is not sufficiently energetic (Kaganovich \& Lyubarsky 2010; Dyks
\& Rudak 2013) \nct{kl10} 
nor flexible enough to explain the observed signal geometry. The induced inverse Compton scattering
(ICS) has been identified as the mechanism that likely reshapes the emitted
radiation (Blandford \&
Scharlemann 1976; Lyubarskii \& Petrova 1996). \nct{bs76,
lp96} However, the scattering has been applied rather to extreme or peculiar phenomena (backward
interpulses, nonbifurcated precursors, giant pulses, Petrova 2004, 2008a,b)
\nct{p2004, p2008, p2008bis} 
and its
role in shaping the regular profile morphology has not been recognized.
%The ICS has not been universally accepted and it is often considered less 
%natural or competitive to the CR. 
In this paper I show that the regular pulsar profiles (main pulses) 
indeed bear a geometric signature of scattering. Other arguments that are
also based on the scattering allow us to make progress with the understanding of bifurcated
emission components, such as those observed in PSR J1012$+$5307 (DRD10).

%geometric properties of pulsar signals indicate that both the
%processes operate in pulsar magnetosphere together.}
%are considered together, numerous pulsar properties can be
%understood and large potential for further understanding is open.

Average profiles of radio pulsars exhibit the mysterious
nested cone morphology with the central component (core) surrounded by 
a pair or two pairs of conal components (Rankin 1983, 1993).\nct{ran83, ran93}
Dipolar $\vec B$ field
offers little structure in the polar regions, 
therefore, since pulsar discovery in 1967, there has been only one 
%dead-end 
interpretation proposed for the size ratio of the
cones (Wright 2003).\nct{wri03} 
The present paper offers alternative view on the origin of the conal morphology 
in Section \ref{morph}.
In the following section some types of profile evolution with frequency are interpreted
in terms of the scattering.

{\sf
}

In Section \ref{bifu} I discuss the origin of bifurcated components (BCs) observed in
radio pulsar profiles. The excessive size of these components, and their
similarity to the frequency-integrated CR microbeam will be both explained in a 
consistent way. Subsection \ref{splitpol} illustrates that the
split part of the CR microbeam has the appropriate polarization to undergo
efficient scattering.
%In the {\bf appendix} I show that the conal
%geometry has to have azimuthal structure, which allows for the bifurcations
%to survive (avoid blurring) even in the conal components of averaged profiles.

%The explanations proposed {\bf are related to} some long-standing problems,
%such as the origin of similar OPM amounts or the origin of subwavelength
%phase lags needed for the strong circular polarization. 
%These subjects are also discussed below.

\section{A model for conal profile morphology} 
\label{morph}

\begin{figure}
\begin{center}
\includegraphics[width=0.47\textwidth]{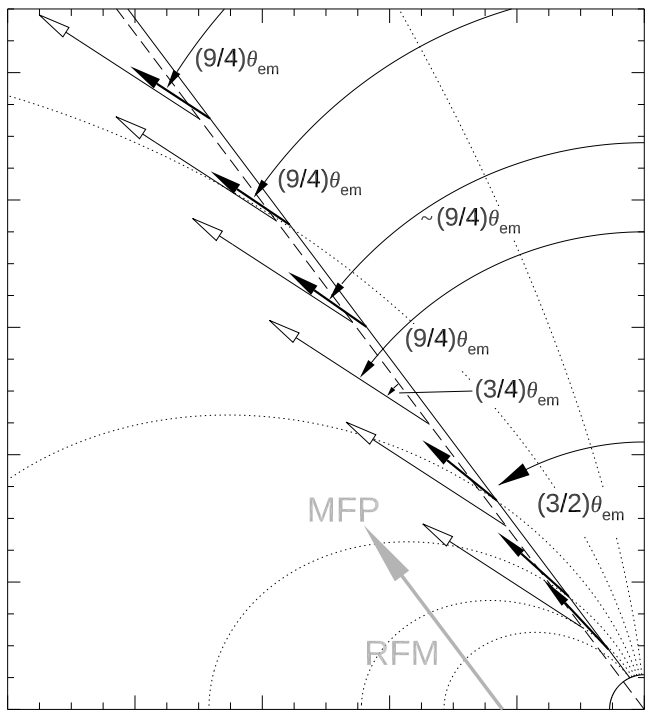}
\end{center}
\caption{Formation of conal emission through scattering of low-altitude
core rays that are emitted at $(3/2)\them$ and propagate upward along the near-diagonal solid line. 
The polar angle of the local $B$-field that 
they encounter (short black arrows) 
is quickly approaching $(9/4)\theta_{\rm em}$. 
For any scattering altitude which is not close to the emission point, the
emitted waves are scattered at this preferred angle, which produces
a cone. 
White-tip arrows show the fixed local direction of $\vec B$ along the dashed radial line which
is parallel to the initial propagation direction. The low-$\nu$ profile
widening is caused by the
increase of the mean free path, as shown with the large grey arrow.
%Short black arrows show the local $\vec B$ direction along the propagation path 
%of the unscattered rays, which form the core.
}
\label{scat}
\end{figure}
\begin{figure}
\begin{center}
\includegraphics[width=0.47\textwidth]{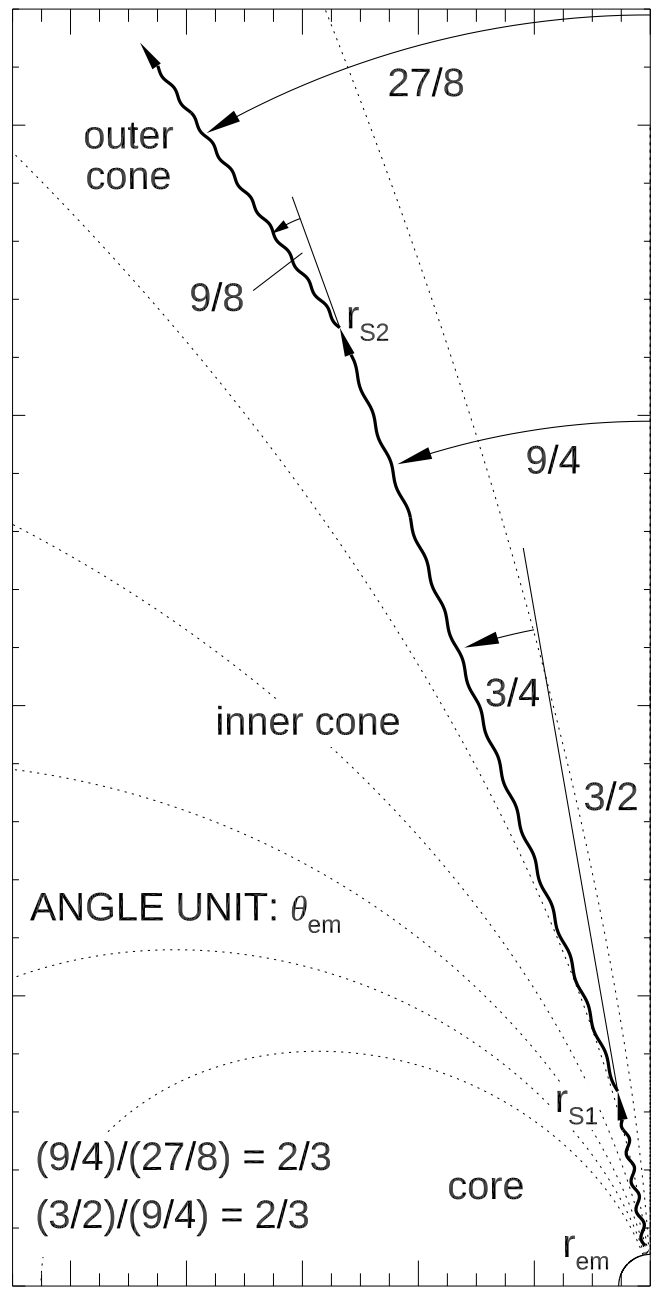}
\end{center} 
\caption{Relations between propagation angles in the core, 
inner cone and the outer cone
rays. The unit of angle is the initial colatitude of an emission point 
$\them$ (arbitrary, but small).  
As explained in Fig.~\ref{scat}, the long mean free path 
and dipolar $\vec B$ field ensure that beams of
consecutive scattering orders are scaled by the factor of $3/2$. 
The inner to outer cone size ratio is thus equal to $0.66$.
}
\label{cones}
\end{figure}

The proposed scenario is based on key features of scattering in the relativistic
limit: 1) the scattered photon is assumed to follow the electron's velocity,
ie.~the local direction of magnetic field; 2) the scattered waves are
blueshifted to a higher frequency, following the rough relation $\nu_{\rm
obs}\sim\gamma^2\nu_{\rm em}$ between the scattered (observed) and emitted
frequency.\footnote{The neglected angular factor is expected to be important for
small angle scatterings (photons following the electron within
$1/\gamma$) which can occur in pulsar magnetosphere, so this simplification 
should be released in a more exact analysis.} I also assume the mean free path is
sufficiently short for the efficient scattering  to occur, where by
efficient it is meant `capable of producing a new pulse component'. 
Estimates of mean free path for nonisotropic scattering are sophisticated even with collective
plasma effects
ignored (eg.~Dermer 1990; Dyks \& Rudak 2000),  \nct{der1990,dr2000}
but they are not necessary to make
the geometric arguments of this paper. Existing estimates show that
efficient scattering in pulsar polar tube is possible in the case of induced
scattering (Blandford \& Scharlemann 1976). \nct{bs76} Assuming that the incident radiation is amplified
to the observed flux densities, Petrova (2008a) \nct{p2008} has shown that detectable
precursors and interpulses can be formed in pulsar polar regions.

In the following it is argued that the cones correspond to the radiation
that is scattered 
in the regime of a long mean free path (MFP), as suggested by the observed
cone size ratio. 
%Assuming that the emitted radiation (that is to be scattered) 
%represents the zeroth order of scattering, one may say that the cones 
%generally correspond to different scattering orders.
Consider a low-altitude core ray emitted tangentially to the local magnetic field,
at the polar angle $(3/2)\them$, where $\them$ is the colatitude of the emission
point, located at a radial distance $r_{\rm em}$. 
The ray propagates along the solid slanted line in Fig.~\ref{scat}. 
In the relativistic limit, the ray will be scattered
along the local $\vec B$ field (black vectors) at a scattering angle
$\theta_{\rm sc}$ 
as measured
between the initial and scattered ray direction.
The dipolar magnetic field along any radial line 
(including the dashed line parallel to the ray in Fig.~\ref{scat})
makes a fixed angle $\theta_x$ with respect to the radial direction, as shown with the white
tip arrows. The angle is equal to half the radial line tilt, in
this case: $\theta_x = (3/4)\them$.  The corresponding polar angle (measured
from the dipole axis) is equal to 
\begin{equation}
%\theta_{\rm obs} = (3/2)\them+(3/4)\them=(9/4)\them,
\theta_{\rm obs} = \frac{3}{2}\them+\frac{3}{4}\them=\frac{9}{4}\them,  
%= \frac{3}{2}\left(\frac{3}{2}\them\right),
\label{sceq}
\end{equation} 
so it is larger by the factor $3/2$ than the radial line
tilt. %  $(9/4)\them=(3/2)(3/2)\them= $  
%Direction of $\vec B$ along the dashed radial line (parallel to the ray in
%Fig.~\ref{scat}) is shown with the white tip arrows. 
Since the solid ray path is nearly radial far from $\rem$, 
the propagating ray is cutting the local $\vec B$ field at an increasing angle 
which quickly approaches $\theta_x$. 
%(the local direction of $\vec B$ is shown with black vectors).
%$(3/2)\them$ 
Only near the emission point, ie.~for a short MFP 
(hereafter denoted $\esc$), % $\esc\ll r_{\rm em}$, 
the scattering direction depends on $\esc$ or altitude. 
For nonlocal scatterings ($\esc\gg r_{\rm em}$)  all the rays 
are scattered at the same polar angle $(9/4)\them$. 
This can be interpreted as the inner cone.\footnote{Or the outer cone, if the emitted radiation
is assumed to form a ring associated with the last open field
lines. I that case, however, the core is an extra feature that needs
explanation.}
%see below.}
The width ratio between the emitted and scattered beam is
\begin{equation}
%$R_{io}=(3/2)/(9/4)=2/3$. 
R_{io}=\frac{3}{2}\them\left(\frac{9}{4}\them\right)^{-1}=\frac{2}{3}. 
\end{equation}

\begin{figure}
\begin{center}
\includegraphics[width=0.47\textwidth]{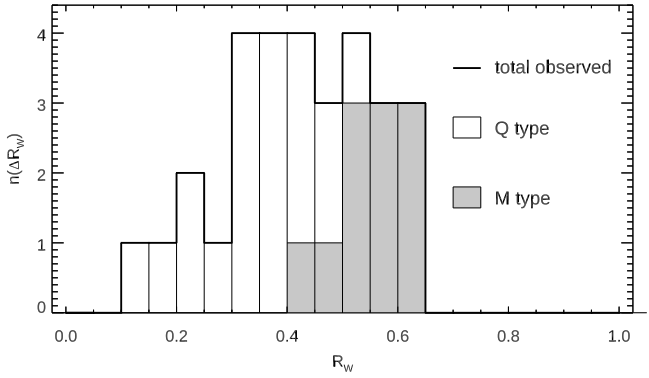}
\end{center}
\caption{Histogram of component separation ratio ($R_W=W_{\rm in}/W_{\rm
out}$) for the inner and outer conal pair in quadruple and multiple profiles.
After Dyks \& Pierbattista (2015). 
}
\label{psepdistr}
\end{figure}

\begin{figure*}
\begin{center}
\includegraphics[width=0.8\textwidth]{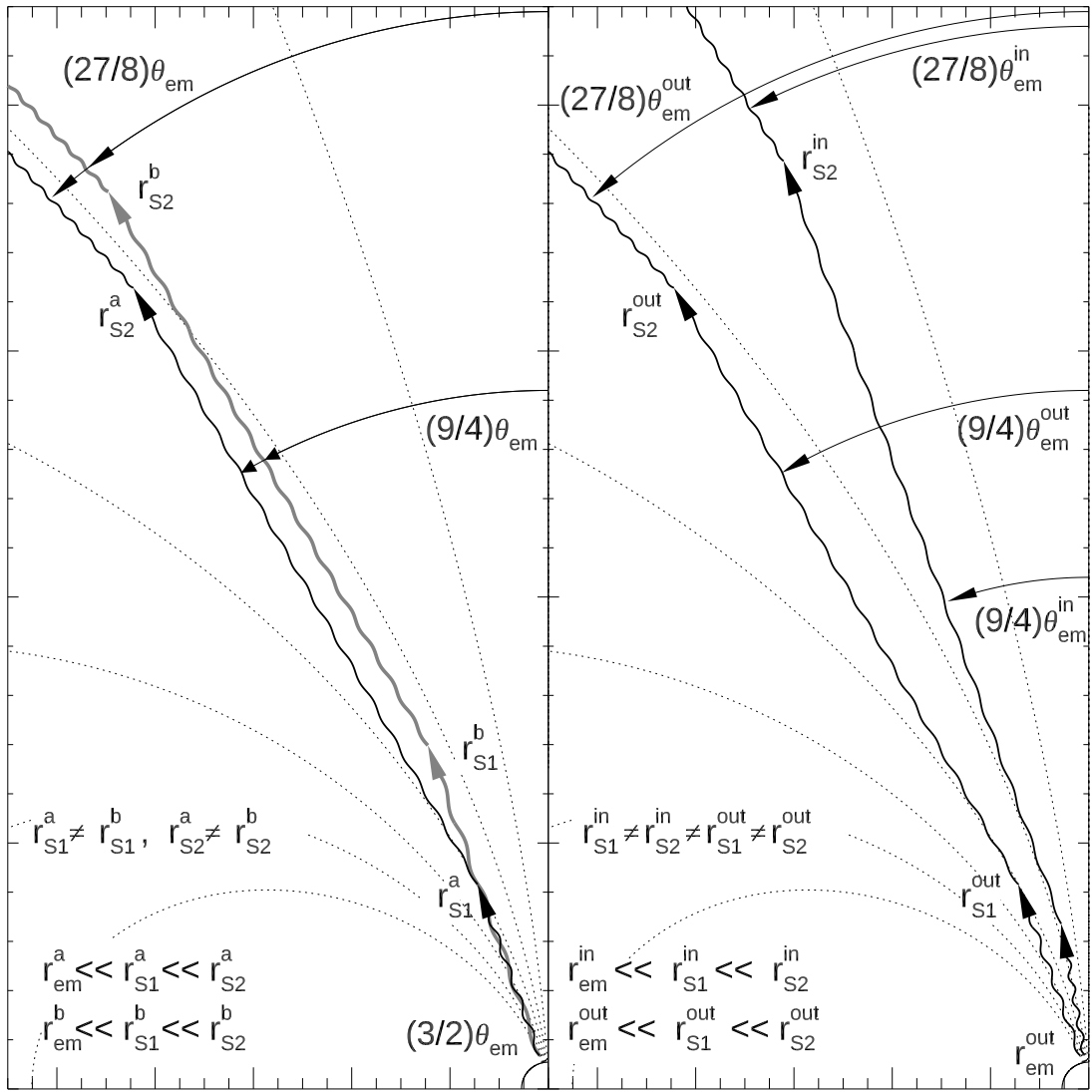}
\end{center}
\caption{Geometry of long-MFP scattering for a pair of rays.  
Left: Two rays, `a' (black) and `b' (grey), emitted 
at the same polar angle $(3/2)\them$. In spite of different scattering altitudes, 
the scattered rays of a given order
propagate at the same polar angles, as shown with the double-tip arrows. 
Right: Two rays (inner and outer) emitted at
different polar angles $(3/2)\theta_{\rm em}^{\rm in}$ and $(3/2)\theta_{\rm em}^{\rm
out}$. In each scattering order, the propagation directions of scattered rays
preserve the initial angle ratio $\theta_{\rm em}^{\rm in}/\theta_{\rm em}^{\rm
out}$. 
}
\label{come}
\end{figure*}

The above reasoning is recursive, so if the conditions allow for
the second-order scatterings at long MFP, the outer cone is produced at the angle of
$(3/2)(9/4)\them=(27/8)\them$ (Fig.~\ref{cones}). 
The ratio of the inner to outer cone width is again 
$R_{\rm io}=2/3=0.66$. 

The observed cone size ratio has been estimated in a nearly model-independent way
from the measurements of the component separations in Q and M
profiles (Dyks \& Pierbattista 2015, hereafter DP15). 
%In that method, the peak-to-peak separations $\Delta_{\rm in}$ and
%$\Delta_{\rm out}$ are measured for the inner and outer pair of conal
%components and the ratio $R_W=\Delta_{\rm in}/\Delta_{\rm out}$ is
In that method, the peak-to-peak separations $W_{\rm in}$ and
$W_{\rm out}$ are measured for the inner and outer pair of conal
components and the ratio $R_W=W_{\rm in}/W_{\rm out}$ is
calculated. The value of $R_W$ only depends on the distance from the beam centre (called impact angle) 
at which the sightline is cutting the beam.  
The upper limit in the $R_W$ distribution corresponds to the central sightline traverse 
through the beam (see fig.~1 in DP15\nct{dp15}) in which case we have
$R_W^{\rm max}=R_{io}$. As shown in 
Fig.~\ref{psepdistr} %fig.~4c
(after DP15\nct{dp15}, see also table 1 therein) ten percent of all the Q and M pulsars have the highest 
$R_W$ value of $0.63$ which is in good agreement with the scattering origin. The number of
pulsars with near-maximum $R_W$ is much lower than predicted for a conal structure,
however, this can result from the beam suppression in the central-traverse
region, as observed for the interpulse of PSR J1906+0746 (Desvignes et al.~2019). \nct{dkl19} 

Another known estimate of observed $R_{\rm io}$ is 
larger ($\sim\negthinspace\negthinspace\negthinspace0.75$, Rankin 1993;
Mitra \& Deshpande 1999; Kramer et al.~1994),\nct{ran93, md99, kwj94}
 however, 
it was derived with indirect methods that use the curve of polarization angle 
(PA). As argued below (Section \ref{frevol}), %(Section \ref{cohtra}), 
the curve may be distorted by the scattering itself, and by other effects 
(eg.~distortions that can be modelled as the coherent superposition of
polarization modes, Dyks, Weltevrede \& Ilie~2021). \nct{dwi21} 

Dipolar magnetic field geometry implies that for long $\esc$,  
the scattering angle does not depend on scattering radii 
(Fig.~\ref{come}, left) which must facilitate the appearance of the cones. 
Since all 
scattering angles are proportional to $\them$, essentially the cones are 
reflected version of the core (Fig.~\ref{come}, right). This is in
line with the usually-similar width scale of observed core and cone components. 

In the above-described version, the model assumes multiple scatterings. 
In a case with only a single scattering, a version of the model is possible in which
the inner cone is formed by rays emitted near the last open lines (the emission
is considered as the zeroth scattering order). 
The core in triple profiles can then be interpreted by a sightline grazing
this 
inner (emitted) cone, whereas the scattered cone forms the single pair of conal components. 
Quadruple profiles (with four components) can also be interpreted in this
model.
%, with the inner conal components corresponding to the emitted cone, whereas the outer
%components to the scattered cone.
%Then the rays produced by the single (first) scattering must be identified as the outer cone. 
However, in multiple-type profiles (with five components) the observed core component
then requires a separate justification, see also Sect.~\ref{splitpol}.

\section{Frequency evolution of profiles}
\label{frevol}

\begin{figure}
\begin{center}
\includegraphics[width=0.47\textwidth]{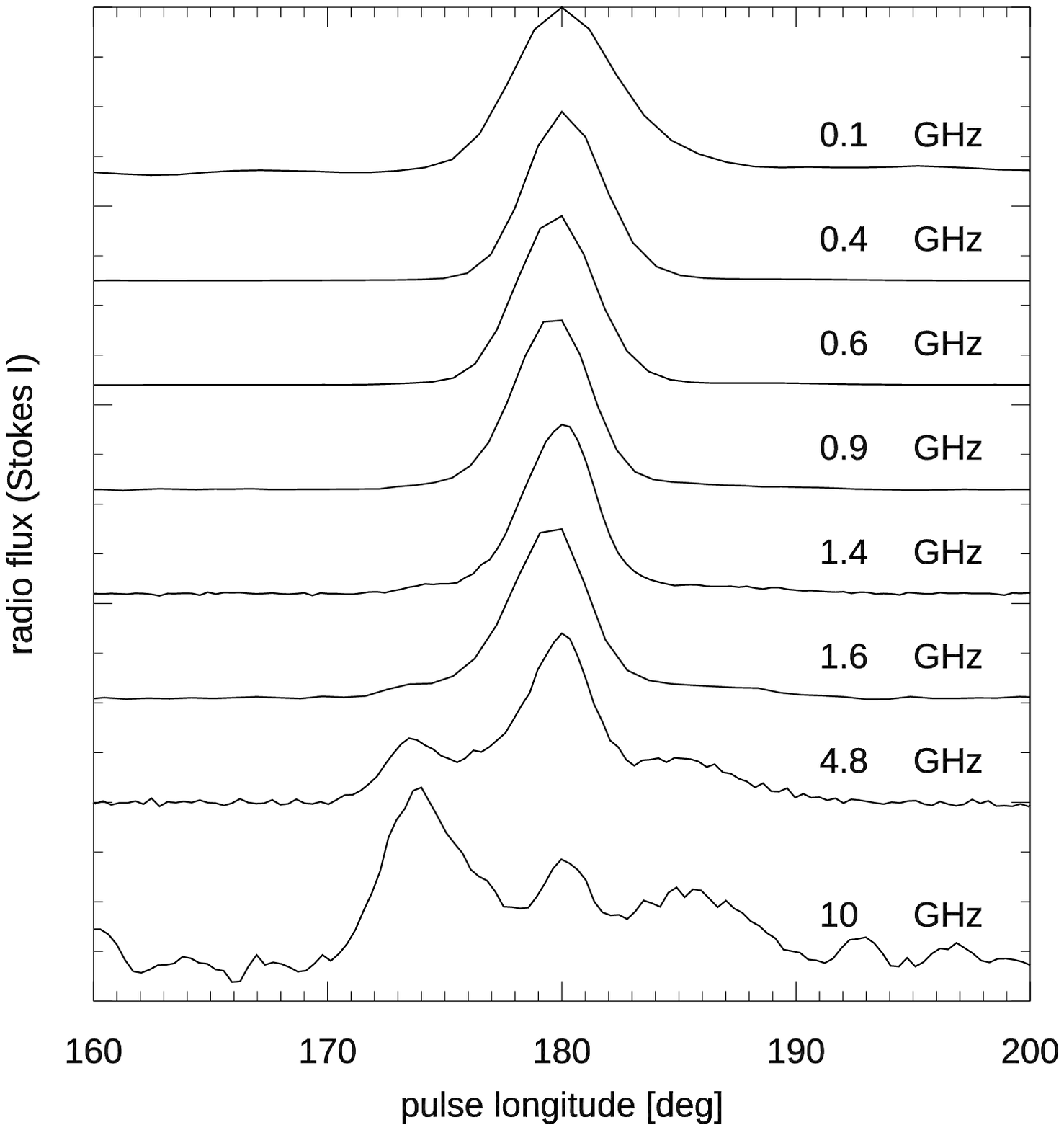}
\end{center}
\caption{Average profiles of B1642$-$03 at eight frequencies that increase
towards the figure's bottom. Note the emergence of conal components above 1 GHz. Data from EPN
%(Kassim \& Lazio 1999; 
(Gould \& Lyne 1998; Seiradakis et al.~1995; von
Hoensbroech \& Xilouris 1997).
}
\label{b16}
\end{figure}
\nocite{gl98, sgg95, hx97}

The scattering model is in line with some types of observed frequency
evolution of profiles. In particular, it is supported by profiles in which conal components emerge at high
frequencies, as in PSR B1642$-$03 (Fig.~\ref{b16} with frequency increasing
downwards). This is consistent with the blueshifted nature of the scattered conal
emission. 

%\subsection{Interpretation of a radio pulse profile in frequency and longitude}

\begin{figure}
\begin{center}
\includegraphics[width=0.39\textwidth, height=0.8\textheight]{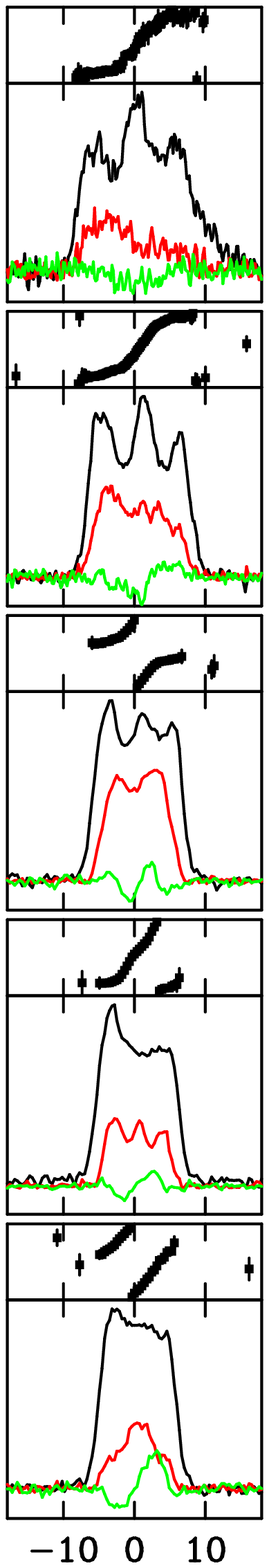}
\end{center}
\caption{Frequency evolution of the polarized profile of PSR B1700-32 (fig.~4 in
Johnston et al.~2008). %\nct{jkm2008} 
From top to bottom, the frequencies
are: $243$ and $322$ MHz (GMRT) and $0.69$, $1.4$ and $3.1$ GHz (Parkes Telescope). 
%The PA at the Parkes frequencies is calibrated in absolute way. 
The range of PA axis is $(-90^\circ, 90^\circ)$, the
longitude is in degrees. Note the stairs-shaped PA at $1.4$ GHz, and the emergence of conals with flat PA.
}
\label{simo}
\end{figure}
\begin{figure}
\begin{center}
\includegraphics[width=0.39\textwidth, height=0.7\textheight]{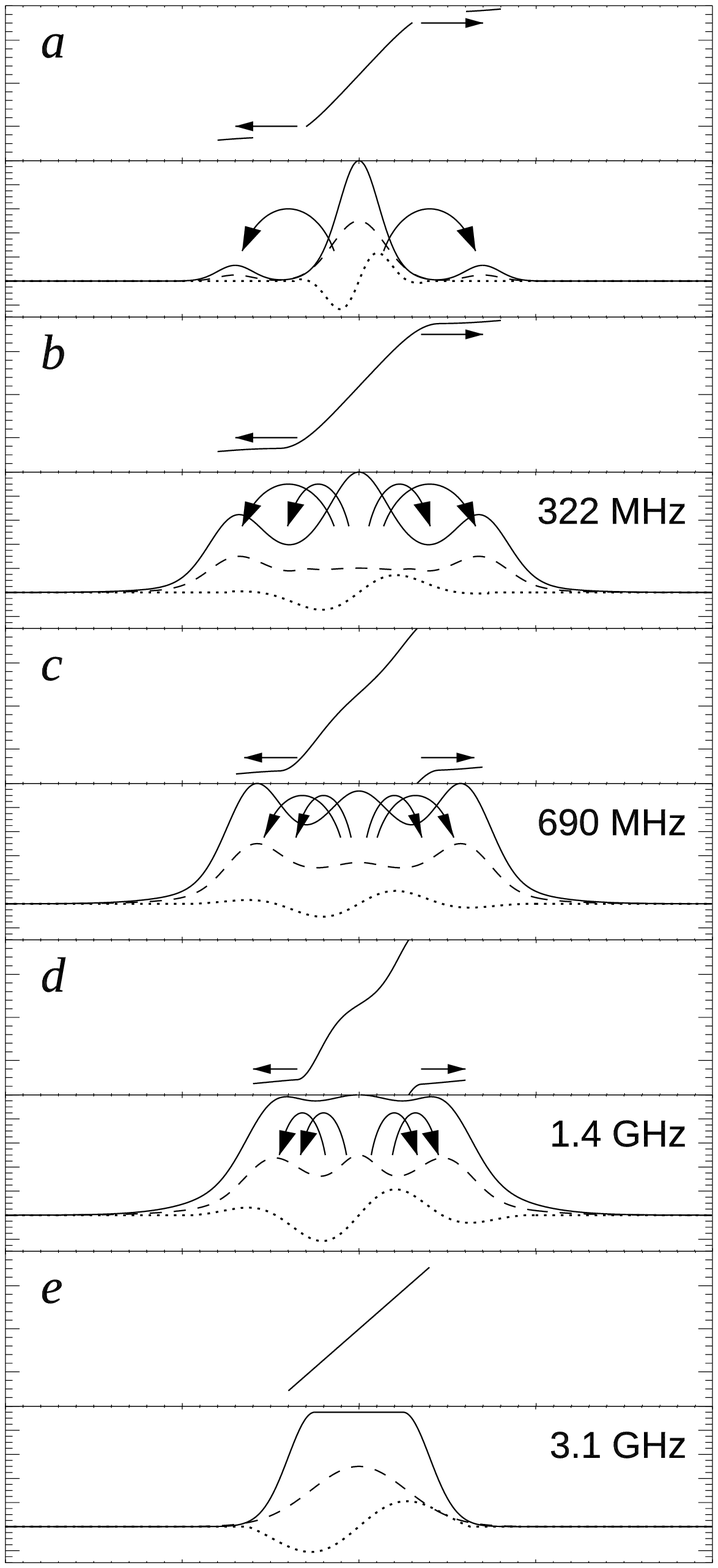}
\end{center}
\caption{Schematic interpretation of the frequency evolution 
observed in PSR B1700-32 (Fig.~\ref{simo}).  
Bent arrows show the ray displacement by the
scattering. Horizontal arrows show the associated PA displacement, which may
lead to the flattening of the peripheric 
PA curves. Top two panels reflect the behaviour of B1642$-$03 (Fig.~\ref{b16}).
%For more details see text.
}
\label{nuevol}
\end{figure}

On the other hand, if the conditions are favorable, it is possible to observe well developed 
core and cones within the same frequency band. This is the case 
when the spectrum of the incident radiation (to be scattered) extends to
sufficiently low frequency, or the blueshift is small on account of geometric
effects.  
Fig.~\ref{simo} shows the $\nu$ evolution for PSR B1700$-$32, as observed by
Parkes Telescope and GMRT (Johnston et al.~2008).\nct{jkm2008} 
To interpret such profile through scattering  
one must assume that the MFP 
decreases with increasing frequency (this assumption is the basis of the
interpretation shown in Fig.~\ref{nuevol}).  
The upper part of Fig.~\ref{simo}
(GMRT, $243$ and $322$ MHz) suggests that below $243$ MHz the profile is dominated by the
core and likely follows the evolution which is observed in PSR B1642$-$03
(Fig.~\ref{b16}, cf.~Fig.~\ref{nuevol}a,b).\footnote{According to this view, the
profiles of Figs.~\ref{b16} and \ref{simo} represent different parts of a
single scattering sequence shown in Fig.~\ref{nuevol}. The former set of
profiles (for B1642$-$03) is continued in B1700$-$32.}
The blueshifted conals in Fig.~\ref{nuevol}a
emerge at the unique distance from the core that
corresponds to the long-MFP scattering angle (eq.~\ref{sceq}).
With increasing $\nu_{\rm obs}$ (and shorter MFP) the angles of scattering
become smaller, because at lower altitudes the emitted photons
cross the local $\vec B$ at smaller angles (black arrows in
Fig.~\ref{scat}). This makes the profile narrower.  
The triple form becomes less pronounced since the short MFP
scatterings do not share the common scattering angle (which is only the
feature of long-MFP scatterings). The pulse window is thus filled in with radio
flux (Fig.~\ref{nuevol}b,c). 

%The main discussed effects that
%shape this `beautiful example of a triple profile' are shown in
%Fig.~\ref{nuevol}. The lowest-$\nu$ stage (panel a) is not observed in
%B1700-32, but is often seen in other pulsars:
%the sinusoid-like $V$ under the narrow core is
%consistent with the convolution of CR microbeams (Michel 1991, p.~356) \nct{m91} 
%while the central PA curve may reflect the rotating vector model (RVM). 
%With increasing $\nu$ the profile
%developes weak separated conal components, ie.~$\esc$ is large but
%decreasing.  
%At 322 MHz (panel b)  
%scatterings at a smaller angle fill in the beam. 

As shown in Fig.~\ref{nuevol} (horizontal arrows), it is reasonable to expect that 
the transport of radio flux towards the profile 
periphery makes the PA curve flat. The outer parts thus do not have to correspond to the
rotating vector model (RVM). 
The peripheric PA may
consists of the `interior' PA values that have been relocated outwards by the
scattering.\footnote{This intramagnetospheric PA flattening is different
from the interstellar effect discussed in Karastergiou (2009).} \nct{k2009} 
This effect is not accounted for in the usual PA modelling, which may
explain the notorious problems with the RVM fitting of
flat outer parts of PA curves (eg.~PSR B1857$-$26, Mitra \& Rankin 2008). \nct{mr2008}   
%At 690 MHz the ICS MFP is shorter for most rays, which makes the profile
%narrower (the rays are less defocused by the scattering, see black arrows in
%the bottom-right corner of Fig.~\ref{scat}). 

Since the scattering rate is different for different polarization modes, the
cones are likely to have orthogonal polarization to that in the core, as it is
often observed (eg.~Srostlik \& Rankin 2005). \nct{sr2005} 
It is also the case for B1700$-$32: at $1.4$ GHz the PA curve assumes a stairs-like shape, 
caused by the predominance of the other polarization mode in the core region. 
%The total polarisation degree does not drop at the modal transitions, 
Despite the dominant OPMs swap in longitude, there is no trace of this in
total polarization degree, which may be
interpreted as a mostly-coherent orthogonal polarization mode jump (Dyks et
al.~2021). \nct{dwi21}
%%which appears because at this $\nu$ the phase lag $\dxo$ between the total O and X modes is closer to the quarter-wave value. 
%Thus, at this $\nu$, with the increasing longitude $\Phi$   
%the observed polarization state rotates non-equatorially on the Poincare sphere. 

\subsubsection{The radius-to-frequency mapping}

The scattering geometry in dipolar $\vec B$ field (Fig.~\ref{scat}) implies small scattering angles
for short MFP.  The small profile width at $3.1$ GHz (bottom panels in Figs.~\ref{simo} and \ref{nuevol}) 
then corresponds to the local low-altitude scatterings that occur in the bottom
right corner of Fig.~\ref{scat}. 
Thus, according to this model, the radius-to-frequency mapping (RFM) is caused by the changes of the
scattering MFP. At
high $\nu_{\rm obs}$ the short % $\esc$ 
MFP is erasing the preference of the $1.5\theta$ direction and fills in the polar tube, % at the scattering altitude, 
thus producing the narrow boxy profiles.

Pulsar profiles then seem to be shaped by the
scattering-driven ray reflection.
% {\bf with possible  interference, which may
%lead to} the $\nu$-dependent mode alternation ($\ma$) (see Fig.~\ref{taxo}). 
The cones in such model are produced at
a higher altitude than the core, and the outer cones are located higher than
the inner cones. This is what has been found observationally in the studies based on aberration-retardation
effects (Gangadhara \& Gupta 2001; Dyks et al.~2004; Krzeszowski et al.~2009).
\nct{gg01, drh04, kmg09}
The core may also have different radio spectrum, because
it must be mostly made of unscattered rays: since the scattering widens the
incident beams, the narrowest beam likely corresponds to the unscattered
(incident) radiation. 

As discussed in Section \ref{splitpol}, the core likely consists of the central
(filled-in) part of the CR microbeam(s), which has the unfavourable polarization orientation of
the extraordinary mode, thus avoiding the scattering. Unlike that, the split O-mode part of the beam is
widened into cones (deflected outwards by scatterings). 
When at high frequency the scattering ceases to deflect the O-mode outwards, 
both polarization modes are observed together within the
narrow core-like boxy profile. This implies depolarization which is indeed 
often observed at high $\nu$ \nct{nsk15}
(eg.~Noutsos et al.~2015; also see Fig.~\ref{simo} in this paper).\footnote{Refraction would imply similar
mode-dependent effects, however, refraction does not imply specific cone size ratio.
Still, I do not exclude a possible contribution of refraction at the
lowest observed frequencies (few hundreds of MHz).}

%and, possibly, rays produced by small-angle scatterings.

%The sinusoid $V$ profile is limited to the narrow core interval at low
%$\nu$, but it extends over wider $\Phi$ interval at higher $\nu$, to streach
%basically across the entire boxy profiles at the highest $\nu$. 
%This `sine-widening effect' is clearly visible in PSR B1857$-$26 (fig.~6 of JKMG08)
%and can be understood in similar way as the peripheric RVM  
%The outward redirection of the core rays is transporting
%causes the sinusoid $V$ profile to
%extend across a wider interval of pulse longitude
%within the boxy 

\section{The nature of bifurcated components} 
\label{bifu}

\begin{figure*}
\begin{center}
\includegraphics[width=0.87\textwidth]{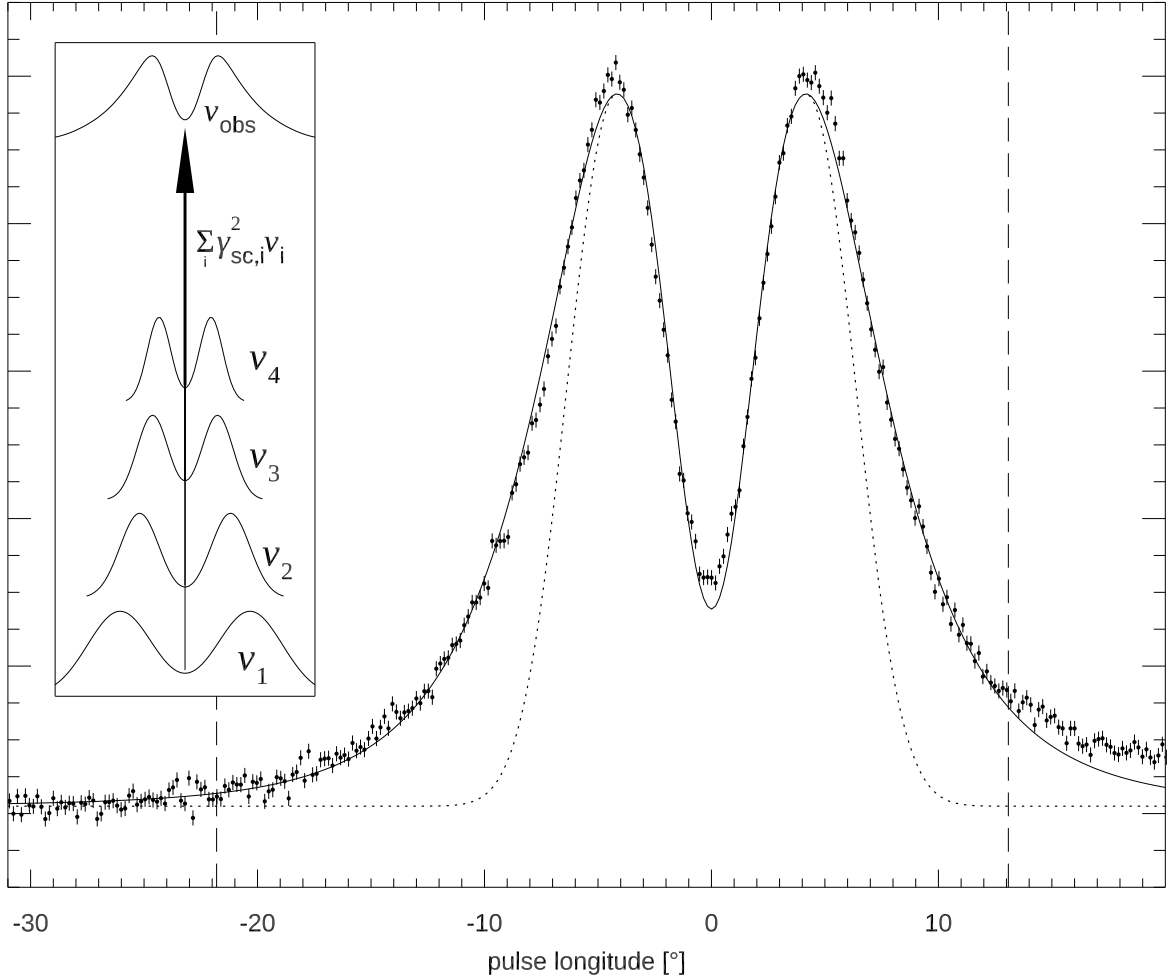}
\end{center}
\caption{The BC in PSR J1012$+$5307, as observed at $0.82$ GHz by GBT (from Dyks et
al.~2010). 
In spite of the narrow
bandwidth of $64$ MHz ($8\%$) the BC is well modelled by the $\nu$-integrated
O-mode-dominated CR beam (within the dashed verticals $\chi^2/{\rm
dof}=3.6$, error bars are statistical $1\sigma$, vertical scale in arbitrary
units). Dotted line shows the steep outer wings in the
frequency-resolved CR beam.  
Inset: the bright outer wings are formed by
Doppler-stacking (through blueshifting) of wide low-$\nu$ CR microbeams: waves of
different 
$\nu$ are scattered by electrons with different energy $\gsc$.  
The large size of the low-$\nu$ microbeams is preserved by the beam-copying 
scattering (see next figure).
}
\nct{drd10} 
\label{bc}
\end{figure*}

\begin{figure*}\begin{center}
\includegraphics[angle=90, width=0.67\textwidth]{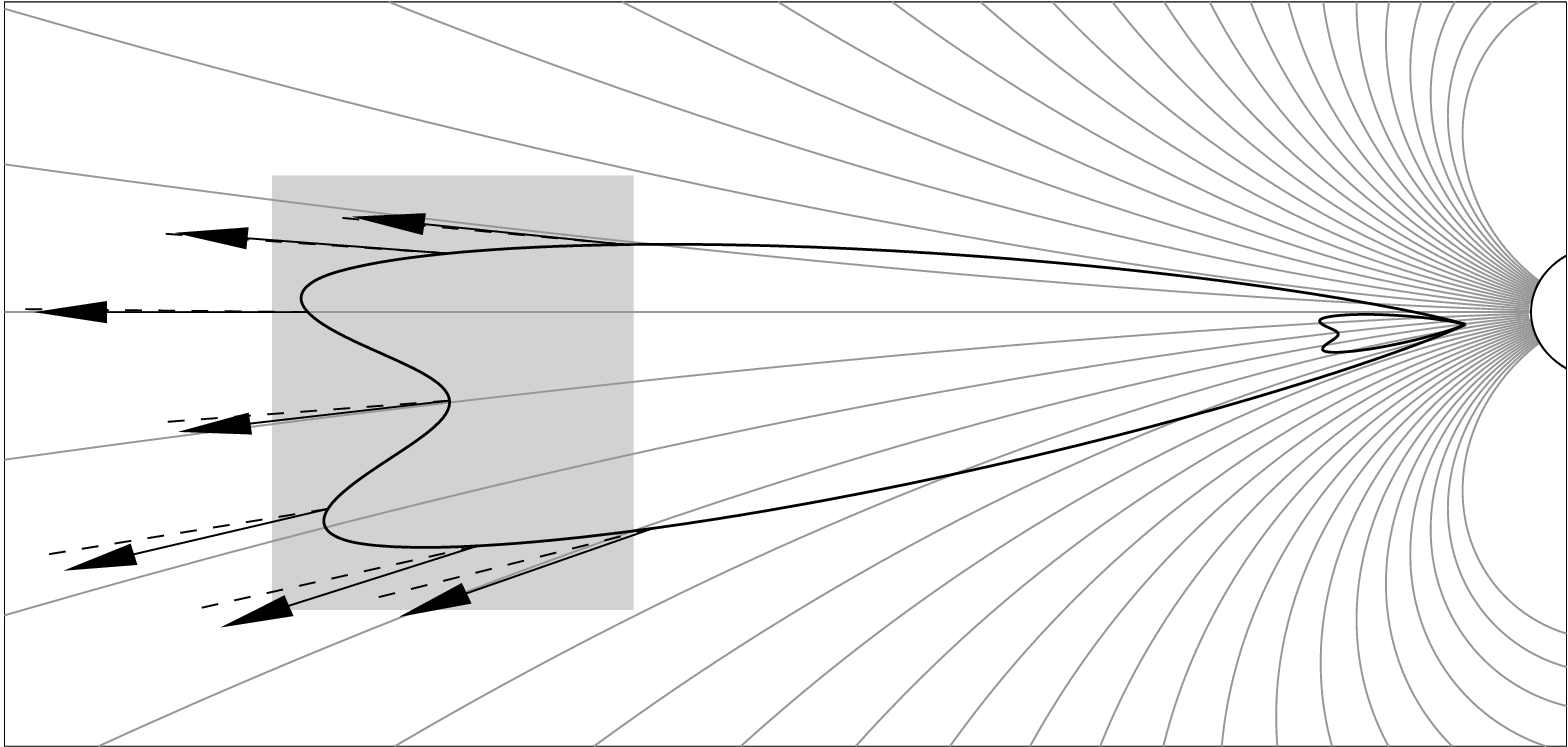}
\end{center}
\caption{The mechanism of the ray-to-ray beam copying by scattering. Individual rays in the emitted beam
(propagating %in the page plane 
along the dashed lines) are
locally scattered along the black arrows that represent the local velocity direction for distant electrons. 
The wide low-$\nu$ beam shape is thus copied to the high $\nu_{\rm obs}$ with its large width
approximately preserved. 
%The dashed rays are radial with respect to the emission point on the right.  Two wavefronts are shown. 
Unlike the B-field lines, the dashed rays are located within the
page plane. The rays cross the local $\vec B$ at an angle larger than $1/\gamma_{\rm sc}$ because $\vec B$
within the grey rectangle is penetrating the page at some angle 
($\sim\theta_x$), so that the side view
resembles Fig.~\ref{scat}.}
%\footnote{The angle between the dashed rays and the arrows of the
%local $\vec B$ is larger than $1/\gamma_{\rm sc}$ because they are
%misaligned in the plane orthogonal to the page. When viewed from a side the
%rays cross $\vec B$ in the way shown in Fig.~\ref{scat} (imagine that
%the dashed rays in Fig.~\ref{magni} are in the page plane, whereas $\vec B$
%within the grey rectangle is piercing the page at some angle.}}
\label{magni}\end{figure*}

%\begin{figure*} \begin{center}
%\includegraphics[angle=90, width=0.77\textwidth]{magni.ps} \end{center}
%\caption{The mechanism of the ray-to-ray beam copying by scattering. Individual rays in the emitted beam
%(propagating along the dashed lines) are
%locally scattered in the direction of black arrows that represent the local velocity direction for distant electrons. 
%The low-$\nu$ beam shape is thus copied to high $\nu$ with its large width preserved or even slightly enlarged. The 
%sections of dashed line are radial with respect to the emission point at bottom. Two wavefronts are shown.}
%\label{magni} \end{figure*}

The bifurcated components (BCs) can be divided in two classes: 
narrow conal components that merge quickly with increasing frequency
($\Delta\propto\nu^{-1/2}$, see fig.~6 in Dyks et al.~2007) \nct{drr07}  
and do not resemble the CR microbeam shape. A typical example is the trailing
conal component in the millisecond PSR J0437-4715 (Navarro et al.~1997; Dai
et al.~2015; Oslowski et al.~2014) \nct{nms97, dhm15, ovb14} although
conal BCs have also been observed in normal pulsars (PSR B1946+35, Mitra \&
Rankin (2017), B1933+16 Mitra et al.~2016).\nct{mr2017,
mra2016}\footnote{The bifurcations are illustrated, but not mentioned in the cited papers.}     
The other BCs are very
wide precursors -- the primary example is the strong and very symmetric interpulse
precursor in PSR J1012$+$5307 (Fig.~\ref{bc}, based on fig.~8 in DRD10). \nct{drd10} 
The peaks in this BC merge with $\nu$ at the rate
$\Delta\propto\nu^{-0.35}$ which is close to that of the CR microbeam 
($\Delta\propto\nu^{-1/3}$). The BC has a shape that is similar to the 
$\nu$-integrated CR microbeam. 

It is suggested here that the two BC types correspond to two different scattering regimes. The narrow
fast-merging
BCs are scattered within a magnetospheric region with small spread of velocity
directions: $\Delta\hat v \ll 1/\gsc$, where the hat means a unit vector, and
$\gsc$ is the Lorentz factor of the scattering electrons. 
With no widening by the velocity spread, this implies the
scattered beam size $\Delta\sim 1/\gsc$. 
In the case of ICS, the observed frequency is roughly equal to
$\nuobs\approx\gsc^2\nuem$, where $\nuem$ represents the peak frequency of the emitted
CR spectrum. Hence $\Delta\propto(\nuem/\nuobs)^{1/2}$, ie.~the
$\nu$-dependence of these BCs is directly consistent with the ICS origin (Dyks et
al.~2007). \nct{drr07}

The strong BC in PSR J1012$+$5307 merges with $\nu$ at the rate expected for CR. However, 
although it has been observed within a narrow frequency band of $8\%$, 
it has a shape that is very similar to the $\nu$-integrated CR microbeam
(Fig.~\ref{bc}). 
%In the $\nu$-resolved microbeam the intensity of CR drops steeply in the periphery
%(at a super-Gaussian rate\footnote{Dotted line in Fig.~\ref{bc} follows eq.~4 in
%DRD10, see also Section \ref{splitpol} below.}), %
In the $\nu$-resolved microbeam the intensity of CR drops steeply in the periphery
(at a super-Gaussian rate, ie.~the dotted line in Fig.~\ref{bc} follows eq.~4 in
DRD10), 
whereas the $\nu$-integrated microbeam has well-developed wings, which
extend far outwards (eq.~8 in DRD10).\footnote{The analytical formulae for
the shapes of both $\nu$-resolved and $\nu$-integrated beams are provided in Section
\ref{splitpol}.}  
% (see fig.~3 in Dyks et al.~2010). \nct{drd10} 
The observed $\nu$-resolved BC should therefore be strongly dissimilar to
the $\nu$-integrated CR microbeam that is fitted in Fig.~\ref{bc} (solid
line). 
% It is thus the success of the fit in
%Fig.~\ref{crb}, which is problematic, not the fit's slight disagreement
%Contrary to this expectation, the Gaussians do not fit the BC (see fig.~3 in 
%Dyks et al.~2010),\nct{drd10}  
Instead, the $\nu$-integrated CR beam produces a decent fit: 
between the dashed vertical lines $\chi^2/{\rm dof}=3.6$ (DRD10).\nct{drd10}    
The strong outer wings cannot be explained by spatial convolution,
because it would mostly fill in the central minimum in the observed
BC (fig.~3 in Dyks \& Rudak 2013). \nct{dr2013} 
Moreover, the BC is about ten times wider than the CR
microbeam, which has the peak separation of $\Delta_{\rm cr} \approx
0.8^\circ/(\nu^{1/3}\rho_{B,7}^{1/3}\sin\delta_{\rm cut}\sin\zeta)$, where
$\rho_{B}=10^7 {\rm  cm} \rho_{B,7}$ is the curvature radius of electron trajectory, $\delta_{\rm cut}$
is the angle at which the sightline crosses the split fan beam, and $\zeta$ is 
the viewing angle between the sightline and the pulsar rotation axis (see fig.~2
in Dyks et al.~2012). \nct{dr12}   

\subsection{Doppler magnification}

The two problems (the large width and the forbidden similarity of the $\nu$-resolved
to the $\nu$-integrated beam) become solved and consistent with each other as soon as the
BC originates from scattering in the regime of large spread of velocity
directions ($\Delta\hat v \gg 1/\gsc$) and non-negligible spread of scattering
electron energies ($\gsc$). If the wide low-$\nu$ CR microbeam is scattered in 
a region with diverging $\vec B$ field 
(grey rectangle in Fig.~\ref{magni}), then each ray is directed along 
the black vectors of the local $\vec B$-field. Although the aberration works 
as usual in this scattering (sending a ray along the electron velocity), 
the entire original CR microbeam is not collimated
(relativistically beamed) because different parts of the CR microbeam
(different rays) are scattered at
different magnetospheric locations by electrons moving in different directions 
(nonlocal scattering).      The beam size and 
shape is thus copied (ray by ray) from the low $\nu$ to the high observed frequency
$\nuobs\approx\gsc^2\nu$. 
The spread of $\vec B$ can slightly magnify the beam (by less than $3/2$, as
shown in Fig.~\ref{magni}),
however, the key magnification comes from the translation in the
frequency space. % The magnification occurs in the frequency space: 
The beam
essentially does not become wider -- it is just made wider than expected at the observed
$\nuobs$, but roughly of the same size as at the much lower $\nu$ before the scattering 
(see inset in Fig.~\ref{bc}). 

\subsection{Spectral compression}
\label{compr}

The second property -- similarity to the $\nu$-integrated microbeam, results
from the spread of the scattering electron energy. Following the
gamma-square rule 
$\nuobs\approx\gsc^2\nu$, higher-energy electrons pick up lower
frequencies from the emitted CR spectrum, and deposit the flux at the same $\nuobs$.
The lower-$\nu$ CR is emitted in wider microbeams, which contribute predominantly to the 
outer wings of the observed high-$\nu$ BC. All the different CR microbeams in the emitted
CR spectrum join the narrow observed band, and make the
BC look $\nu$-integrated.  Thus, the $\nu$-integrated shape results from the
convolution of the electron energy distribution with the emitted CR
spectrum (the high-$\nuobs$ profile in the inset of Fig.~\ref{bc} is the sum 
of the four inset profiles shown below).

\subsection{Electron energies and energy budget}

The BC of J1012$+$5307 has the $1$-GHz width of about $8^\circ$ but 
it may be enlarged by oblique sightline cut (small $\delta_{\rm
cut}$) by unknown factor. Assuming $0.1$ rad for its real  width, the electrons
have the Lorentz
factor of $\gamma_{\rm em}\sim10$, and emit very low-frequency 
CR at $\nucr\approx 7\ {\rm GHz}\ \gem^3/(\rho_B [\rm cm])\sim 1$ MHz. To reach the
observed GHz band, the scattering electrons must have $\gsc\sim30$. This is
pretty close to $\gamma_{\rm em}\sim10$, which suggests the same energy
distribution for both the emitting and scattering electrons. 
The mechanism can thus be called the curvature-self-Compton radiation. 

The ten-fold increase of beam size implies that the energy budget
problem %(Dyks \& Rudak 2013; Gil \& Melikidze 2010) \nct{dr2013, gm2010} 
is abated by three orders of magnitude on account of the increased volume alone.  
Furthermore, the energy available within the radio band is no longer restricted by the
spectral properties of the CR, since the observed BC is mostly powered by the
scattering electrons. 
In the case of the BC of J1012$+$5307 
the required efficiency of energy transfer to the radio band has been previously estimated
as $10^{-4}$ of the maximum power of a plasma stream, as limited by
the electric potential drop and possible stream width (Dyks \& Rudak 2013). \nct{dr2013} 
Since this estimate is proportional to the stream crossection $A$ (see eq.~3
therein), the tenfold increase of microbeam size decreases this power
transfer efficiency to $10^{-6}$. 
%The identification of the BC as Doppler-shifted and
%ray-to-ray-copied CR microbeam, thus solves all the energy-related problems. 

\subsection{Does the split beam have the right polarization to be scattered?}
\label{splitpol}

Blandford and Scharlemann (1976) have shown that only the ordinary polarization mode is 
 efficiently scattered in strong magnetic field. The mode is polarized in
the plane that contains the wave vector $\vec k$ and the local magnetic field
$\vec B$. The split part of the curvature radiation beam is known to be polarized across 
the plane of a B-field line. Superficially, this may lead us to
erroneously think that the split part of the beam may not be scattered, which is not the case. 
This point 
%is not strongly emphasized in classical textbooks, and it 
is sometimes overlooked, so it will be discussed
explicitly in the following.

Fig.~\ref{beampol} schematically shows the curvature radiation beam 
emitted by an electron moving along the curved magnetic field line (dashed
line).
The customary derivation of the beam's intensity results in the
well known formula consisting of two terms:
\begin{equation}
I_{\rm cr} = f_1 + f_2\sin^2\psi
\end{equation}
where $\psi$ is the angle between $\vec k$ and $\vec B$.\footnote{The form of $f_1$ and $f_2$ is 
given below in eq.~(\ref{ncrll}) but it is not important for the present discussion.}   
The first term represents the filled-in part of the beam (grey lobe in
Fig.~\ref{beampol}), which is polarized along the electron acceleration vector
$\vec a$ (the polarization direction is shown with the short double-tip arrows). 
The second term (psi-squared term) corresponds to the split part of the beam 
(black solid lobes in Fig.~\ref{beampol}) which is  
polarized along $\vec k\times \vec a$, where $\vec k$ is the wave vector
pointing towards the observer. The angle $\psi$ 
is exagerrated in the figure, because even at a very low frequency ($\sim1$ MHz) the 
CR beam has the opening angle around $0.1$ rad. Therefore, the
polarization of the split beam is approximately orthogonal to the B-field
line plane. However, in spite of such almost orthogonal orientation, the
polarization is within the $\vec k$-$\vec B$ plane.
%parallel to the projection of $\vec B$ on a plane orthogonal to the line of sight. 
Thus, it is the split part of the curvature beam that
has the polarization of the ordinary mode.\footnote{Whereas the central filled-in part,
which is polarized parallel to the B-line plane, is polarized in the direction
of the X mode.} Therefore, contrary to the initial pesimistic suspicion, the split fan beam of
the curvature radiation has the polarization favorable for the 
scattering.\footnote{The polarization of the beam parts with respect to the B-field line
plane (parallel/orthogonal) is described by adjectives that must be replaced
with each other as soon as the reference to the propagation modes is made
(parallel/orthogonal beam parts correspond to orthogonal/parallel
polarization of modes). In DRD10 \nct{drd10} 
the B-line-based polarization of beam parts has been directly extended to
the names of the polarization modes 
which is incorrect (hence the names `X mode' and `O mode' should be
replaced with each other, and similarly the symbols $\parallel$ and 
$\perp$). The mode-related designation of intensity terms (indices
$\parallel$ and $\perp$) is used eg.~in Rybicki \&
Lightman \nct{rl79} (1979, eq.~6.29), whereas Jackson (1975) \nct{jac1975} uses the indexing based
on the B-field line plane.  }   

\subsubsection{The question of the core component}

With the X mode polarization, the central microbeam part (or a spatially
convolved bundle of such beam parts) is likely to represent the undeflected (not
widened) central part of the profile, ie.~the core. The core component has
indeed been shown to have the low-altitude opening angle of the polar tube (Rankin
1990).

\subsection{Wings of the classical curvature radiation beam}

\begin{figure}
\begin{center}
\includegraphics[width=0.47\textwidth]{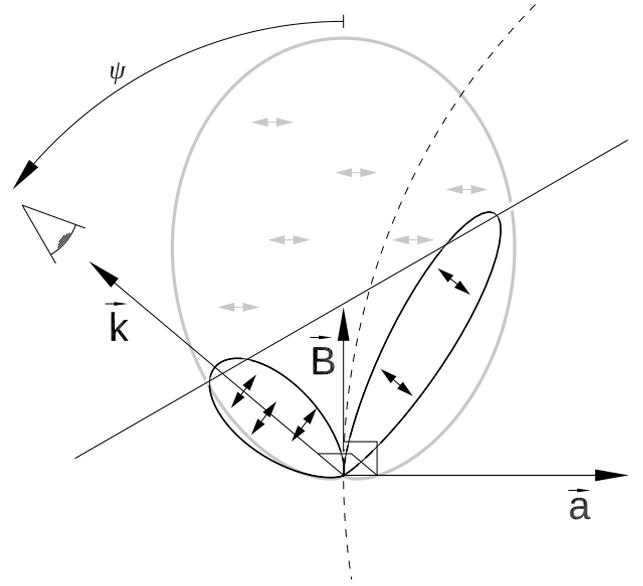}
\end{center}
\caption{Customary decomposition of the curvature radiation beam in two
orthogonally polarized parts: the part peaking along $\vec B$ (grey) is
polarized along the acceleration $\vec a$ whereas the bifurcated part (black 
solid) is polarized orthogonally to the plane containing $\vec k$
and $\vec a$. The latter polarization is in the plane of $\vec k$ and $\vec
B$ which is the polarization of the ordinary mode. The dashed line is the
curved electron trajectory.}
\label{beampol}
\end{figure}

Following the remarks of Section \ref{splitpol}, we may write down the CR beam formula using the
indices $\perp$ and $\parallel$ with reference to the $\vec k\times \vec a$ plane (hence appropriate to
the modes). The energy emitted per unit solid angle and unit frequency is:
\begin{eqnarray}
\eta_{\rm cr} & = & \eta_\perp + \eta_\parallel = \\
 & = & \frac{q^2\omega^2}{3\pi^2c} \left( \frac{\rho}{c} \right)^2\left[
\xi^2 K^2_{2/3}(y) + \xi K^2_{1/3}(y)\sin^2\psi \right],\label{ncrll}
\end{eqnarray}
where
\begin{eqnarray}
\xi &=& 1/\gamma^2 + \psi^2, \ \rm and\\
y&=&\frac{\omega\rho}{3c}\xi^{3/2}.
\end{eqnarray}
Here $\rho$ is the radius of curvature of electron trajectory,
$\omega=2\pi\nu$, $c$ is the speed of light, and $K$'s are the modified
Bessel functions. The angle $\psi$ is measured from the plane of the
electron trajectory  (Fig.~\ref{beampol}). 

Using the lowest term in the large-argument expansion of the modified Bessel functions 
\citep{as1972}:
%\citep{Abramowitz \& Stegun (1972): \nct{as1972}
\begin{equation}
K_{2/3}(y) \simeq K_{1/3}(y) \simeq \left(\frac{\pi}{2y}\right)^{1/2} e^{-y}
\end{equation}
the shape of the outer wings ($\psi > 1/\gamma$) becomes:
\begin{equation}
\eta_{\rm cr} {\rm (wings)} \simeq \frac{q^2\nu\rho}{c^2}\ |\psi| \
e^{-|\psi|^3/\sigma_{cr}^3} \left( 1  +
\frac{\sin^2\psi}{\psi^2} \right) \label{res}
\end{equation}
where we introduce a parameter $\sigma_{cr}=(3c/(2\omega\rho))^{1/3}$.
Ignoring the difference between $\sin\psi$ and $\psi$ it is found that wings
of both modes have equal strength and the same `super-Gaussian' shape of
type $|\psi| \exp{(-a |\psi|^3)}$. The wings
are shown as dotted line in Fig.~\ref{bc}).

In the $\nu$-integrated case the CR beam represents the power emitted per steradian:
\begin{eqnarray}
I_{\rm cr} &=& I_\perp + I_\parallel = \frac{7q^2c}{256\pi\rho^2} \left(
\frac{2\beta^7}{\cos\psi} \right)^{1/2} \times \\
& \times & (1 - \beta\cos\psi)^{-5/2}\left(1 + \frac{5}{14}
\frac{\beta\cos\psi\sin^2\psi}{1-\beta\cos\psi}\right), \label{int} 
\end{eqnarray}
where $\beta$ is the electron velocity in units of $c$. Since
$1-\beta\cos\psi\simeq \psi^2/2$, both modes have wings of type
$(1-\beta\cos\psi)^{-5/2}\propto\psi^{-5}$ (cf.~eq.~12 in DRD10). 

Depending on how large part of the CR spectrum has been integrated within the
observed bandwidth (or has been stacked there by the scattering) the shape
of a component may be closer to the limit (\ref{res}) or (\ref{int}).
Additionally, a convolution with the spatial emissivity distribution may be
needed.\footnote{This discussion assumes that the beam is not deformed by
the process that makes it coherent.} Some contribution from the spatial extent likely affects the BC in
Fig.~\ref{bc}. The central flux in the model shown with the solid line is fully
provided by the filled-in X-mode part of the beam (normalized first term of
eq.~\ref{int}) which likely plays the role of the spatial extent. 

\section{Discussion}

\subsection{Implications for subpulse modulations}

The properties of bifurcated components imply that spectral effects
play crucial role in determining the observed flux of single pulse radio emission. 
First, the scattering must blueshift the emitted radio spectrum to reach the
high frequency band normally used by radio telescopes. Second, the
flux is further amplified by compressing the emitted spectrum into the narrow
bandwidth used in the observations. From the observational point of view, where flux
modulations by a factor of a few make a noticeable difference, the scattering 
looks like a crucial flux-determining factor.

%According to Blandford \& Scharlemann (1976) \nct{bs76} the scattering `almost
%certainly' cannot be the coherency mechanism, since scattering does not produce photons, contrary
%to masers. Indeed, for the scattering process to start operating the optical path for
%scattering must be sufficiently high, which, according to the literature (eg. Petrova ... refs), 
%requires the initial CR emission to be already amplified. However, from the observational point of view, where flux
%modulations by a factor of a few make noticeable difference, the scattering is crucial, 
%since it can move the spectrum out of view, or compress the full radio spectrum into the telescope bandwidth. 

When the modulation phenomena are attributed to the spectral
variations, it seems natural that many of them do not involve displacement of
subpulses in pulse longitude. It appears that the temporal variations of
the electron energy spectrum (hence of the observed radio flux and spectrum) 
are at least as important for the flux modulations as the lateral rotation from the ExB
drift.
%}  

\subsection{Cones or fan beams?}

\begin{figure}
\begin{center}
\includegraphics[width=0.47\textwidth]{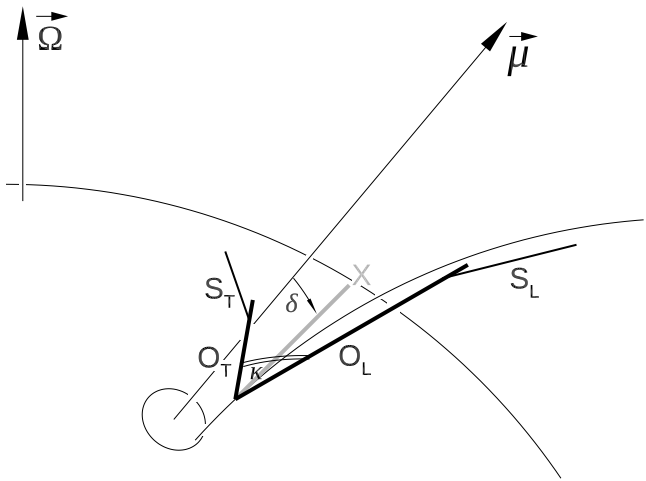}
\end{center}
\caption{Geometry of the low-altitude fan-beam emission from a bent plasma stream 
(the arc emerging from the polar cap). 
Sections $\rm O_T$, $\rm X$, and $\rm O_L$ represent the axes of the CR beam parts
shown in Fig.~\ref{beampol}. When the beam's opening angle $\kappa$ is larger than the beam's tilt $\delta$, 
the O-mode rays $\rm O_T$ and
$\rm O_L$ roughly follow the geometry of Fig.~\ref{scat} with the scattered rays
$\rm S_T$ and $\rm S_L$ following similar beam size ratio $W_{\rm O}/W_{\rm
S}\approx2/3$.}
\label{branch}
\end{figure}

The presented model is clearly ambivalent with respect to the conal-or-fan beam
alternative.

Section 1 suggests the concentric rings (nested conal beams) whereas the BCs are interpreted
as cuts of sightline through the split-fan beams. It does not help to say that conal
components are different than the bifurcated precursors observed  in J1012$+$5307,
because the bifurcations are also observed in the normal conal components of the pulsars
mentioned at the beginning of Section \ref{bifu}. The bifurcations have been considered as the main justification for
the nonconal beam geometry (system of radial fan beams, see fig.~18 in DRD10),
\nct{drd10} because 
the diverging plasma flow in the polar region is expected to smear any structure in
magnetic latitude (fig.~1 in Dyks 2017a).  \nct{dyk17} 
One possibility is to claim that conal components in different pulsars may have
different origins:
they are either deflected (scattered) rays from the ring-shaped emission regions, 
or they result from a cut through the fan beam emission
from a bent `conal' stream (a stream lying not in the main meridian). 

Another option is to assume that only the fan beams exist (instead of rings or cones). 
The system of fan beams is known to exhibit the same RFM and
aberration-retardation effects as the conal region (Karastergiou \& Johnston 
2007; Dyks et al.~2015) \nct{kj07, dr15} and has
favorable statistical properties (Lyne \& Manchester 1988; Chen \& Wang 2014). 
\nct{lm88, cw14} 
By allowing for the wide microbeams, one may also consider 
a version of the fan beam model with the pulse structure formed by the microbeam shape,
instead by the flow structure.   
In such microbeam-based fan beam model the scattering can produce a doubly-split fan beam 
as shown in Fig.~\ref{branch}
(with the core formed by the unscattered central X mode). The four-component
profile of J0631$+$1036 may be interpreted as 
a doubly-split (four-lobed) fan beam structure on account of problems
described in section 3 of Teixeira et al.~(2016). \nct{trw16}
In fact, by assuming that the microbeam opening angle $\kappa$ is larger than the beam's tilt
$\delta$ with respect to the dipole axis, it is possible to derive the same 
`conal' size ratio of $2/3$ (Fig.~\ref{branch}). This is because for
$\delta\sim0$, the O-mode rays 
$\rm O_T$ and $\rm O_L$  propagate essentially in the same way as shown in
Fig.~\ref{scat}. However, the ratio of $2/3$ seems to be a lower
limit in the case of such microbeam-based doubly-split fan beam, which
then likely does not reproduce the distribution of
Fig.~\ref{psepdistr}.\footnote{Moreover, if a conal component comes from
the scattering of a single O-mode lobe (say $\rm O_T$ into $\rm S_T$) we must again explain 
why $\rm S_T$ would be sometimes observed as bifurcated. 
One would have to argue that the scattered rays $\rm S_T$ and
$\rm S_L$ activate CR emission of new streams which themselves emit
bifurcated microbeams. This would lead to a branching stream with causally
connected branches and interesting
modulation properties.} 
For this reason it is the mixed scenario (involving the cones and fan beams)  
that appears to be the most likely possibility.
Whereas the scatterings can explain the
apparent conal properties, they do not necessarily prove that the geometry
behind all conal components is conal.
%can be seen by considering the other limiting case: for $\delta=90^\circ$
%the microbam is immersed in the radial B-field so the rays are scattered towards 
%the interior of the profile ($W_S \lt W_O$). Initially, ie.~at low altitudes
%we have $W_S \gt W_O$ (Fig.~\ref{branch}) and with the increasing altitude the O and S rays 
%are expected to be aligned, hence the peak separation ratio $W_O/W_S$  must increase above $2/3$. 

%Since some conal components (both in the millisecond and normal pulsars) are
%bifurcated, they are best explained through the sightline passage across fan
%beams. However, the logic of Section 2 (on cone size ratio) and Section 3
%(on frequency evolution) of this paper was clearly based on the
%conal geometry.

\section{Conclusions} 

It has been shown that the general properties of the inverse Compton scattering 
can explain the apparent nested cone geometry with the observed conal size ratio. 
The blueshift of the scattered radiation is consistent with some types of
frequency evolution of profiles, in particular with the emergence of conal
components at high $\nu$. 
The scattering is also consistent with discreteness of the invoked polar emission
rings (cones), as well as with their altitude structure as derived from
the aberration-retardation effect (higher altitude of outer rings).  

Assuming that the incident waves are the curvature radiation, it is also possible to
understand the peculiar properties of bifurcated components. 
This includes their rate of merging, their flat outer wings, and their oversized
dimensions. The width-saving transportation of the emitted CR beam up to 
a much higher observed frequency must be responsible for the large size of BCs. 
This implies that the relativistic beaming fomula $(\Delta\sim1/\gamma)$ does not
always hold, in the sense that it may not be directly applied to the width
of the Doppler-magnified components in pulsar profiles. Moreover, the low-$\nu$ microbeams (to be
scattered) have angular sizes comparable to the entire polar tube, which
decreases the energy requirements for generation of coherency.

It is also concluded that both the shape of components in average radio pulse
profiles, and the observed signal modulations are strongly affected by
processes that occur in the frequency space. In particular, the outer wings
of components are very sensitive to the effect of
spectral stacking of flux within the observed narrow frequency band. 
The modulation properties depend on whether the scattered spectrum includes
the observed band, which depends on details of scattering geometry and
plasma flow structure.

This paper solves the main obstacles in our previous papers on BCs (the
problem of large width and the $\nu$-integrated shape). It has been
emphasized now that the
orthogonal-to-fieldline polarization of the split part of the curvature beam 
is actually consistent with the orientation of the ordinary mode waves, 
which is parallel to the $\vec k$-$\vec  B$ plane (and this polarization mode is expected to
be scattered efficiently). 
%Judging from the quality of the fit (DRD10), the observed BC of PSR J1012$+$5307 
%is likely affected by minor contribution from spatial convolution effects 
%that operate  in addition to the spectral effects. A more precise modelling must
%therefore involve many parameters (both spatial and spectral), with the risk of possible ambiguity of
%solution. Nevertheless, 
This paper then offers a self consistent scenario
providing further credence to the idea, that resolved and magnified 
views of microscopic radiation patterns are made visible to us by pulsars.

\section*{Acknowledgments} 
I thank Simon Johnston for Fig.~2.
% and Bronek Rudak for comments on the manuscript. 
This work was supported by the grant 
2017/25/B/ST9/00385 of the National Science
Centre, Poland.
\section*{Data Availability Statement} 
This paper is based on published data.
%The data for J1012+5307 are available from the author on request and with permission of Paul Demorest. 
% and with permission of Paul Demorest. 

\bibliographystyle{mn2e}
\bibliography{listofrefs2}

\end{document}